\documentclass[journal]{IEEEtran}
\usepackage{amsmath,amssymb,amsfonts}
\usepackage{graphicx}
\usepackage{textcomp}
\usepackage{xcolor}
\usepackage{cite}
\usepackage{algorithm}
\usepackage{algpseudocode}

\def\BibTeX{{\rm B\kern-.05em{\sc i\kern-.025em b}\kern-.08em
    T\kern-.1667em\lower.7ex\hbox{E}\kern-.125emX}}

\newcommand{\mat}[1]{\boldsymbol{#1}}
\begin{document}
\title{\huge Channel Estimation for Rydberg Atomic Quantum Receivers: Unrolled Phase Retrieval from Holographic Snapshots
\thanks{

	Jian Xiao, Ji Wang and Hongbo Xu are with the Department of Electronics and Information Engineering, College of Physical Science and Technology, Central China Normal University, Wuhan 430079, China (e-mail: jianx@mails.ccnu.edu.cn; jiwang@ccnu.edu.cn; xuhb@ccnu.edu.cn). 
	
	Ming Zeng is with the Department of Electric and Computer Engineering, Laval University, Quebec City, Canada (email: ming.zeng@gel.ulaval.ca).
	
	Xingwang Li is with the School of Physics and Electronic Information Engineering, Henan Polytechnic University, Jiaozuo 454003, China (e-mail: lixingwang@hpu.edu.cn).
	
	 Arumugam Nallanathan is with the School of Electronic Engineering and Computer Science, Queen Mary University of London, E1 4NS London, U.K., and also with the Department of Electronic Engineering, Kyung Hee University, Yongin-si, Gyeonggido 17104, Korea. (email: a.nallanathan@qmul.ac.uk).

	}
	}
\author{Jian Xiao, Ji Wang,~\IEEEmembership{Senior Member,~IEEE}, Ming Zeng, Hongbo Xu,\\ Xingwang Li,~\IEEEmembership{Senior Member,~IEEE}, and Arumugam Nallanathan,~\IEEEmembership{Fellow,~IEEE}}
\maketitle
\begin{abstract}
A model-driven deep learning framework is proposed for channel estimation in Rydberg atomic quantum receivers (RAQRs) based on the measurement of holographic snapshots. Specifically, we develop a Transformer-based unrolling architecture, termed URformer, to solve the non-linear biased phase retrieval problem, which is derived by unrolling a stabilized variant of the expectation-maximization Gerchberg-Saxton (EM-GS) algorithm. Each layer of the proposed URformer incorporates three trainable modules: 1) a learnable filter network that replaces the fixed Bessel kernel in the classic EM-GS algorithm; 2) a trainable gating mechanism that adaptively combines classic updates to ensure training stability; and 3) an efficient channel Transformer module that learns to correct residual errors by capturing non-local channel dependencies. Numerical results demonstrate that the proposed URformer significantly outperforms classic iterative algorithms and conventional black-box neural networks with less pilot overhead.
\end{abstract}


\IEEEpeerreviewmaketitle

\section{Introduction}
\IEEEPARstart{T}{he} demand for ever-increasing data rates and connection density in future 6G networks has spurred research into revolutionary antenna technologies that transcend the limitations of classical radio-frequency (RF) systems \cite{11026007, 10985904}. Concurrently, quantum information technologies are rapidly maturing, offering new paradigms for sensing, computation, and communication \cite{b5}. At the intersection of these fields lie Rydberg atomic quantum receivers (RAQRs) \cite{b4}, an emerging quantum sensor technology that utilizes the extreme sensitivity of highly-excited Rydberg atoms to measure electromagnetic fields with quantum-limited precision \cite{b2}.


Integrating RAQRs into wireless communication systems necessitates a distinct detection architecture that departs from classical RF systems. Specifically, the superheterodyne receiver is the popular architecture in RAQRs \cite{jing2020atomic}, which can exhibit an equivalent linear baseband signal model via a Taylor expansion around the operating point of the strong local oscillator (LO) \cite{b4, 11161431}. However, the superheterodyne RAQRs impose significant hardware complexity for large-scale arrays, as they necessitate dedicated high-bandwidth photodetectors and synchronous demodulation circuitry for each array element to resolve the time-varying intermediate frequency signal. Consequently, in this work, we develop the holographic snapshot-based RAQR architecture to enable scalable parallel readout in RAQRs, which is achieved by utilizing an optical imaging array to capture the spatial intensity profile of the probe laser beam \cite{fan2014subwavelength}, effectively allowing atomic antenna elements to be read out simultaneously in a single exposure. This RAQR architecture also introduces a new signal processing challenge where the magnitude-only measurement is obtained in digital domain and hence results in a non-linear phase retrieval problem, making the task of channel acquisition especially difficult. Foundational solutions to this phase retrieval problem are based on classic iterative algorithms, e.g., the Gerchberg-Saxton (GS) and its expectation-maximization (EM) variants \cite{b2}. Despite their principled design, these methods suffer from two major drawbacks. First, their fixed mathematical structure is rigid and cannot adapt to non-ideal characteristics of a physical RAQR system. Second, they function as local search methods within a highly non-convex optimization landscape, often getting trapped in suboptimal local minima.

To overcome the aforementioned limitations, in this work, we turn to the paradigm of model-driven deep learning for high-accuracy channel estimation in RAQRs, which combines the physical interpretability of iterative algorithms with the data-driven learning power of deep neural networks \cite{9524496}. Specifically, we first formulate the channel estimation problem for RAQRs by deriving the non-linear biased phase retrieval signal model from the underlying quantum measurement principles. Furthermore, we propose a Transformer-enabled deep unrolling network, termed URformer, which transforms the classic iterative algorithm into a trainable framework. This is achieved by systematically replacing the iterative algorithm rigid blocks with three learnable modules: an adaptive neural filter that learns true signal statistics, a dynamic gate that ensures stable convergence, and a bespoke channel Transformer that eliminates non-local residual errors. Numerical results demonstrate that the proposed URformer achieves a state-of-the-art estimation performance compared to traditional iterative algorithms as well as black-box neural networks.

\section{System Model and Problem Formulation}
\begin{figure}[!t]
\centering
\includegraphics[width=3.2in]{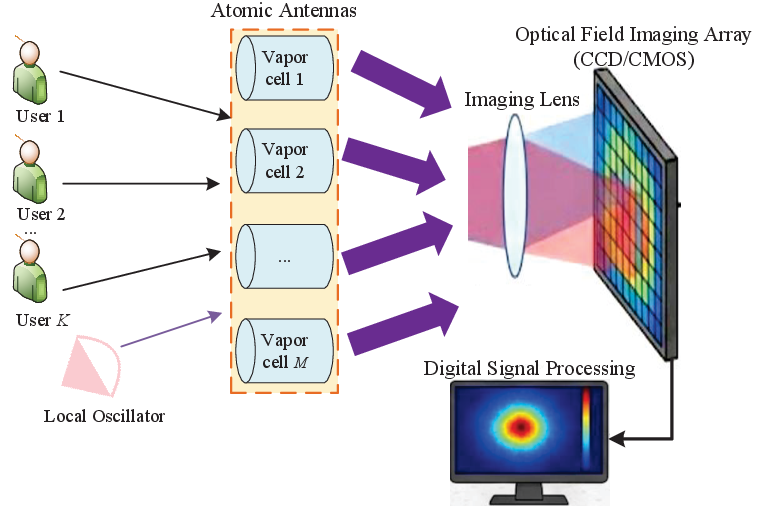}
 \caption{Uplink multi-user systems with the RAQR. By capturing the intensity profile of the probe laser with an optical field imaging array, i.e., holographic snapshots, all spatial channels are detected simultaneously, which reduces the hardware complexity compared to the superheterodyne RAQR.} 
 \vspace{-5mm}
\label{fig1}
\end{figure}
As illustrated in Fig.~\ref{fig1}, we consider an uplink multi-user multiple-input-multiple-output (MIMO) system where $K$ single-antenna users transmit signals to an RAQR equipped with a uniform linear array (ULA) of $M$ Rydberg atomic antennas with inter-antenna spacing $d$. 
\subsubsection{Channel Model}Considering the clustered Saleh-Valenzuela model \cite{6848765}, the channel vector $\mathbf{h}_k \in \mathbb{C}^{M \times 1}$ from the $k$-th user to the RAQR is formulated as a sum over $L$ clusters with each cluster containing $C_l$ rays, which can be expressed as
\begin{equation}
    \tilde{\mathbf{h}}_k = \sqrt{\frac{M}{N_{\text{ray}}}} \sum_{l=1}^{L} \sum_{c=1}^{C_l} g_{l,c}^{(k)} \mathbf{a}(\theta_{l,c}^{(k)}),
\end{equation}
where $N_{\text{ray}}$ is the total number of rays. $g_{l,c}^{(k)}$ and $\theta_{l,c}^{(k)}$ are the complex path gain and direction of arrival (DoA) of the $c$-th ray in the $l$-th cluster for user $k$, respectively. The array steering vector $\mathbf{a}(\theta_{l,c}^{(k)})$ for each ray is given by
\begin{equation}
    \mathbf{a}(\theta_{l,c}^{(k)}) = \left[1, e^{j\frac{2\pi}{\lambda}d \sin \theta_{l,c}^{(k)}}, \dots, e^{j\frac{2\pi}{\lambda}(M-1)d \sin \theta_{l,c}^{(k)}}\right]^T.
    \label{eq:steering_vector_sv}
\end{equation}

The overall MIMO channel matrix is $\mathbf{H} = [\mathbf{h}_1, \dots, \mathbf{h}_K] \in \mathbb{C}^{M \times K}$. During the channel estimation phase, users transmit a known pilot matrix $\mathbf{S} = [\mathbf{s}_1, \dots, \mathbf{s}_P]^T \in \mathbb{C}^{P \times K}$ over $P$ time slots. The complex envelope of the signal field vector incident upon the $M$ antennas at pilot slot $p$ is given by
\begin{equation}
    \mathbf{y}_{p} = \mathbf{H} \mathbf{s}_p,
    \label{eq:signal_field}
\end{equation}
where $\mathbf{s}_p \in \mathbb{C}^{K \times 1}$ is the pilot vector transmitted at slot $p$.

\subsubsection{Holographic Snapshot Measurement Model}
The RAQR operates based on a four-level atomic energy scheme ($|1\rangle \to |2\rangle \to |3\rangle \leftrightarrow |4\rangle$) [8], [11]. A reference LO field vector $\mathbf{b} \in \mathbb{C}^{M \times 1}$ interferes with the impinging multi-user signal field $\mathbf{y}_p$ within the vapor cell. The total RF electric field vector across the array is the coherent superposition:
 \begin{equation}
 \mathbf{E}_{\text{total}} = \mathbf{y}_p + \mathbf{b}.
     \label{eq:1}
 \end{equation}
To achieve scalable parallel readout, we employ an optical field imaging array, e.g., a high-resolution CMOS or CCD camera,  to capture the spatial intensity profile of the probe laser beam. 
By operating on the linear slope of the {electromagnetically induced transparency (EIT)} resonance, the imaging array performs a frequency-to-intensity mapping, where the local Autler-Townes (AT) frequency splitting $\Delta f$ is linearly translated into an intensity variation recorded by the sensor pixels. Specifically, the observed frequency splitting $\Delta f_m$ at the $m$-th pixel antenna is given by
\begin{equation}
    \Delta f_m = \frac{\lambda_c}{\lambda_p} \frac{\Omega_{R,m}}{2\pi},
    \label{eq:at_splitting}
\end{equation}
where $\lambda_c$ and $\lambda_p$ are the wavelengths of the coupling and probe lasers, respectively. $\Omega_{R,m}$ denotes the Rabi frequency which is directly linked to the magnitude of the total electric field acting on the atoms:
\begin{equation}
    \Omega_{R,m} = \frac{|\mu_{34} E_{\text{total}, m}|}{\hbar},
    \label{eq:rabi_freq}
\end{equation}
where $\mu_{34}$ is the electric dipole moment of the Rydberg transition and $\hbar$ is the reduced Planck constant.

We define the measurement vector $\mathbf{z}_p$ at pilot slot $p$ as a holographic snapshot, representing the static magnitude of the spatial interference pattern. Combining Eqs. \eqref{eq:1}-\eqref{eq:rabi_freq}, the array-level measurement model is formulated as
\begin{equation}
    \mathbf{z}_p = \frac{\lambda_c \mu_{34}}{2\pi \hbar \lambda_p} \left| \mathbf{H} \mathbf{s}_p + \mathbf{b} + \mathbf{w}_p \right|,
    \label{eq:measurement_vector}
\end{equation}
where the noise vector $\mathbf{w}_p \sim \mathcal{CN}(\mathbf{0}, \sigma^2 \mathbf{I})$ accounts for the intrinsic system noise, which is dominated by photon shot noise and atomic blackbody radiation, along with the sensor readout noise inherent to the imaging array. Let $\mathcal{G}$ denote the atomic transduction gain $\mathcal{G}=\frac{\lambda_c \mu_{34}}{2\pi \hbar \lambda_p}$.

\subsubsection{Problem Formulation} The objective is to estimate the channel matrix $\mathbf{H} \in \mathbb{C}^{M \times K}$ from the sequence of magnitude-only holographic snapshots $\mathbf{Z} = [\mathbf{z}_1, \dots, \mathbf{z}_P] \in \mathbb{R}^{M \times P}$. Given the known pilot matrix $\mathbf{S}$ and the LO field $\mathbf{b}$, this constitutes a structured MIMO biased phase retrieval problem:
\begin{equation}
\hat{\mathbf{H}} = \arg\min_{\mathbf{H} \in \mathbb{C}^{M \times K}} \frac{1}{MP} \sum_{p=1}^P \left\| \mathbf{z}_p - \mathcal{G} \left| \mathbf{H}\mathbf{s}_p + \mathbf{b} \right| \right\|_2^2.
\label{eq:optimization_problem}
\end{equation}

A key implication of the non-linear measurement model in Eq. \eqref{eq:optimization_problem} is the breakdown of traditional pilot-based channel estimation strategies. In classical linear MIMO systems, the use of orthogonal pilot sequences allows the receiver to decouple the multi-user channel. This principle fails in RAQRs, where the non-linear magnitude operation $|\cdot|$ is applied to the superposition of all signals of users. Due to the property that $|\sum_k \mathbf{h}_k s_{k,p}| \neq \sum_k |\mathbf{h}_k s_{k,p}|$, the orthogonality of the pilot sequences is lost after the measurement is taken. Consequently, the measurement vector for any given pilot transmission is an inseparable function of all channels of users. This invalidates the user-by-user estimation approach and necessitates a joint estimation of the entire channel $\mathbf{H}$.

\section{Proposed Channel Estimation Framework}
\subsection{Baseline Phase Retrieval Algorithms}
The conventional approach to solving the biased phase retrieval problem in Eq. \eqref{eq:optimization_problem} relies on iterative algorithms \cite{b2}. 
Typically, the GS algorithm starts with an initial channel guess $\hat{\mathbf{H}}^{(0)}$, and iteratively refines it by alternating between estimating the phase of signal from the current channel estimate. A statistically robust extension is the EM-GS algorithm, which is to weight the reconstructed signal not just with the estimated phase, but also by a factor that depends on the local signal-to-noise ratio (SNR) proxy $\boldsymbol{\kappa}$. This weighting is performed by the ratio of modified Bessel functions $R(\boldsymbol{\kappa})$, which acts as a soft filter to suppress noisy components. Given the known LO reference field matrix $\mathbf{B} \in \mathbb{C}^{M \times P}$ by stacking the LO vector $\mathbf{b}$ across $P$ pilot time slots, the update rule in the $t$-th iteration is given by \cite{b2}
\begin{align}
&\mathbf{Y}^{(t-1)} = \hat{\mathbf{H}}^{(t-1)} \mathbf{S}^T + {\mathbf{B}}, \label{eq:emgs_y_complex_revised} \\
&\mathbf{\boldsymbol{\kappa}}^{(t-1)} = \frac{2 \mathbf{Z} {\cdot} |\mathbf{Y}^{(t-1)}|}{\sigma^2}, \label{eq:emgs_kappa} \\
&\mathbf{Y}_{\text{EM-GS}}^{(t-1)} = \mathbf{Z} {\cdot} e^{i\angle\mathbf{Y}^{(t-1)}} {\cdot} R(\mathbf{\boldsymbol{\kappa}}^{(t-1)}) \label{eq:emgs_y_rec_revised}, \\
&\hat{\mathbf{H}}^{(t)} = (\mathbf{Y}_{\text{EM-GS}}^{(t-1)} - \mathbf{B}) (\mathbf{S}^T)^{\dagger} \label{eq:emgs_update_revised}.
\end{align}

The performance of the EM-GS algorithm is fundamentally limited by its core filtering function $R(\boldsymbol{\kappa})$. This fixed filter derived under an ideal Gaussian noise model is prone to error propagation in low-SNR conditions. Moreover, as a local search method navigating a highly non-convex optimization landscape, the EM-GS algorithm is susceptible to getting trapped in local optima, making its performance critically dependent on the initial estimate. 

\subsection{Proposed URformer Architecture}
As illustrated in Fig.~\ref{fig2}, we propose a URformer architecture to solve the non-linear channel estimation problem by unrolling a $T_\text{GS}$-iteration EM-GS algorithm into a $T_\text{UR}$-layer neural network, where the fixed computational blocks in EM-GS algorithm, e.g., $R(\boldsymbol{\kappa})$, are replaced by learnable modules. Note that the number of unrolled layers $T_\text{UR}$ is typically much smaller than the $T_\text{GS}$ iterations required by the classic algorithm, stemming from the data-driven learnable modules. A critical challenge in unrolling the EM-GS algorithm is its potential for instability during training. To address this, we introduce a trainable gating mechanism that allows the network to learn a stable update policy. Each URformer layer takes the channel estimate $\hat{\mathbf{H}}^{(t-1)} (1\le t \le T_\text{UR})$ of the previous layer and produces an updated estimate $\hat{\mathbf{H}}^{(t)}$. The initial channel estimate $\hat{\mathbf{H}}^{(0)}$ is generated as a random complex Gaussian matrix with zero mean and normalized variance. A single URformer layer consists of the following three modules.

\subsubsection{Gated Filtering Module} This module replaces the fixed Bessel function filter with a learnable FilterNet and adds a stabilization gate. First, the signal $\mathbf{Y}^{(t-1)} $ is estimated according to Eq. \eqref{eq:emgs_y_complex_revised}, and then the phase is extracted as $\mathbf{\Phi}^{(t-1)}  = \angle \mathbf{Y}^{(t-1)} $. We carry out a direct GS-style reconstruction as
\begin{equation}
    \mathbf{Y}_{\text{direct}}^{(t-1)}  = \mathbf{Z} {\cdot} e^{i\mat{\Phi}^{(t-1)} }.
    \label{eq:l_emgs_y_direct}
\end{equation}
Simultaneously, the parameter $\mathbf{\boldsymbol{\kappa}}^{(t-1)} $ is computed as in Eq. \eqref{eq:emgs_kappa}. We employ a $\text{FilterNet}_{{\psi}^{(t-1)}} $ with trainable weights ${\psi}^{(t-1)} $ as an element-wise operator to learn a filtering function:
\begin{equation}
    R_{\text{learned}}^{(t-1)}  = \text{FilterNet}_{{\psi}^{(t-1)}} (\mathbf{\boldsymbol{\kappa}}^{(t-1)} ), 
    \label{eq:l_emgs_r_learned}
\end{equation}
where the FilterNet is designed as a compact two-layer multi-layer perceptron (MLP) that maps the scalar SNR proxy $\kappa^{(t-1)} $ to a filtering coefficient in the range $(0, 1)$.

Then, the filtered reconstruction is given by
\begin{equation}
    \mathbf{Y}_{\text{filtered}}^{(t-1)}  = \mathbf{Y}_{\text{direct}}^{(t-1)}  {\cdot} R_{\text{learned}}^{(t-1)} .
    \label{eq:l_emgs_y_filtered}
\end{equation}
A learnable gating parameter $\alpha^{(t-1)} $ combines the two reconstructions, which can be expressed as
\begin{equation}
    \mathbf{Y}_{\text{rec}}^{(t-1)}  = \alpha^{(t-1)}  \mathbf{Y}_{\text{filtered}}^{(t-1)} + (1-\alpha^{(t-1)} ) \mathbf{Y}_{\text{direct}}^{(t-1)} ,
    \label{eq:l_emgs_y_gated}
\end{equation}
where $\alpha^{(t-1)}  = \sigma(g^{(t-1)})$ denotes the sigmoid activation of a trainable parameter $g^{(t)}$.

\subsubsection{Linear Estimation Module} Similar to Eq. (13), a linear estimate of the channel is obtained as
\begin{equation}
    \mathbf{H}_{\text{linear}}^{(t-1)}  = (\mathbf{Y}_{\text{rec}}^{(t-1)}  - \mathbf{B}) (\mathbf{S}^T)^{\dagger}.
    \label{eq:l_emgs_h_linear}
\end{equation}

\subsubsection{Residual Correction Module} To remove the contained residual errors in $\mathbf{H}_{\text{linear}}^{(t-1)} $, we develop a Transformer-based residual correction module, termed $\text{Former}_{{\phi}^{(t-1)}}$ with trainable weights ${\phi}^{(t-1)} $, which can be expressed as
\begin{equation}
    \hat{\mathbf{H}}^{(t)} = \mathbf{H}_{\text{linear}}^{(t-1)}  + \text{Former}_{{\phi}^{(t-1)}} (\mathbf{H}_{\text{linear}}^{(t-1)} ),
    \label{eq:l_emgs_h_out}
\end{equation}

To capture the complex non-local dependencies within the MIMO channels which arise from clustered multipath propagation, we develop a channel Transformer module, i.e., $\text{Former}_{{\phi}^{(t-1)}}$, based on the self-attention mechanism \cite{11018390, b8}. Let the input to the module $\text{Former}_{{\phi}^{(t-1)}}$ be the linear channel estimate $\mathbf{H}_{\text{linear}}^{(t-1)}  \in \mathbb{C}^{M \times K}$. The residual correction module is defined by the following sequence of operations.
\begin{itemize}
    \item \textbf{Tokenization:} First, the complex input matrix is decomposed into its real and imaginary components, denoted by operators $\Re(\cdot)$ and $\Im(\cdot)$, respectively, and reshaped into a sequence of $K$ tokens, i.e., $\mathbf{X}_{\text{in}}^{(t-1)}  = \{\Re(\mathbf{H}_{\text{linear}}^{(t-1)} ), \Im(\mathbf{H}_{\text{linear}}^{(t-1)})\} \in \mathbb{R}^{2 \times M \times K}$ is the initial tensor representation. This tensor is then reshaped into a token sequence $\mathbf{T}_{\text{in}}^{(t-1)} = \text{Reshape}(\mathbf{X}_{\text{in}}^{(t-1)} ) \in \mathbb{R}^{K \times 2M}$.

\item \textbf{Embedding:} The token sequence $\mathbf{T}_{\text{in}}^{(t-1)} $ is projected into a higher-dimensional latent space, and positional information is added. This forms the input to the first Transformer layer $\mathbf{V}^{(0)} \in \mathbb{R}^{K \times d_{\text{model}}}$, which can be expressed as
\begin{equation}
\mathbf{V}^{(0)} = \mathbf{T}_{\text{in}}^{(t-1)} \mathbf{W} + \mathbf{P},
\end{equation}
where $\mathbf{W} \in \mathbb{R}^{2M \times d_{\text{model}}}$ denotes the learnable weight matrix of the input projection layer, and $\mathbf{P} \in \mathbb{R}^{K \times d_{\text{model}}}$ is the learnable positional embedding matrix.

\begin{figure}[!t]
\centering
\includegraphics[width=3.4in]{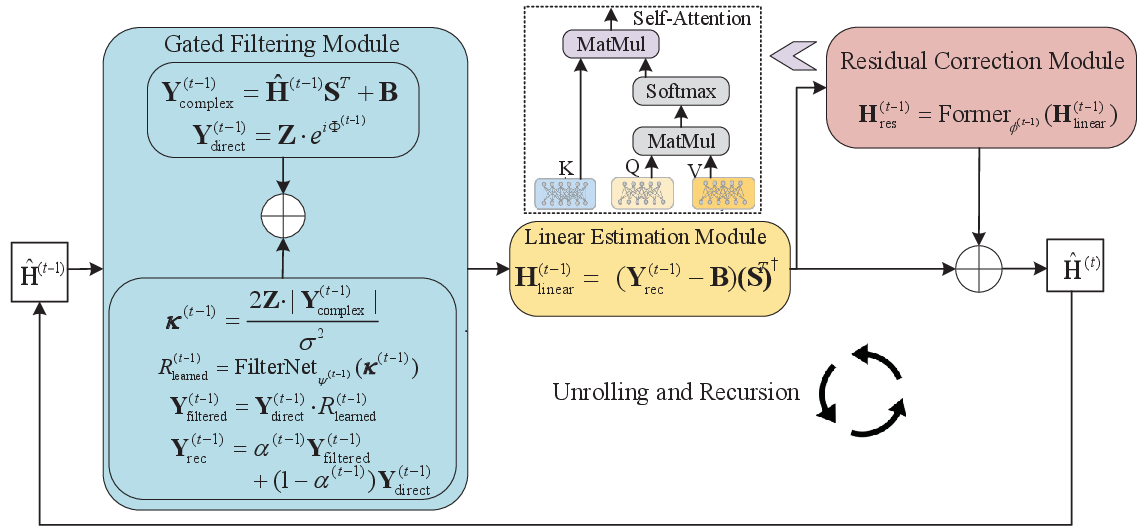}
\caption{Proposed URformer architecture for channel estimation.}
\vspace{-5mm}
\label{fig2}
\end{figure}

\item \textbf{Transformer Encoder Blocks:}
The embedded sequence $\mathbf{V}^{(0)}$ is then processed by a stack of $L_\text{enc}$ identical Transformer encoder layers. For each layer $l=1, \dots, L_\text{enc}$, the multi-head self-attention (MHSA) is carried out, i.e., $\mathbf{A}^{(l)} = \text{MHSA}(\text{LN}(\mathbf{V}^{(l-1)})) + \mathbf{V}^{(l-1)}$. Then, a feed-forward network (FFN) is utilized to obtain the feature tensor $\mathbf{V}^{(l)} = \text{FFN}(\mathbf{A}^{(l)}) + \mathbf{A}^{(l)}$. The output of the final layer is $\mathbf{V}^{(L_\text{enc})} \in \mathbb{R}^{K \times d_{\text{model}}}$. 


\item \textbf{De-Tokenization: } The processed sequence $\mathbf{V}^{(L_\text{enc})}$ is projected back to the original token dimension. The output token sequence is given by $\mathbf{T}_{\text{out}}^{(t-1)}  = \text{FFN}_{\text{out}}(\mathbf{V}^{(L_\text{enc})})\in \mathbb{R}^{K \times 2M}$, which is then reshaped into a tensor $\mathbf{X}_{\text{res}}^{(t-1)}  \in \mathbb{R}^{2 \times M \times K}$. Finally, the two channels of $\mathbf{X}_{\text{res}}^{(t-1)}$ are combined to form the complex residual matrix $\mathbf{H}_{\text{res}}^{(t-1)}=\text{Former}_{{\phi}^{(t-1)}} (\mathbf{H}_{\text{linear}}^{(t-1)} )$. 
\end{itemize}

The $T_\text{UR}$-layer URformer is trained with the normalized mean squared error (NMSE) loss function, which is given by
\begin{equation}
    \mathcal{L} = \mathbb{E}\left[ \frac{\|\mathbf{H} - \hat{\mathbf{H}}\|_F^2}{\|\mathbf{H}\|_F^2} \right].
    \label{eq:loss_function}
\end{equation}
where $\|\cdot\|_F$ denotes the Frobenius norm. 

{The classic GS and EM-GS algorithms have a complexity of $\mathcal{O}(T_{\mathrm{GS}}MKP)$, while the proposed URformer has a complexity of $\mathcal{O}(T_{\mathrm{UR}}(MKP + L_{\mathrm{enc}} K d_{\mathrm{model}}(K + d_{\mathrm{model}})))$. Although the URformer incurs an additional $L_{\mathrm{enc}} K d_{\mathrm{model}}(K + d_{\mathrm{model}})$ term per layer due to the self-attention mechanism, it achieves convergence with significantly fewer unrolled layers ($T_{\mathrm{UR}} \ll T_{\mathrm{GS}}$), thereby reducing the overall computational burden.}

\section{Numerical Results}
In this work, we consider a Cesium (Cs) vapor cell-based RAQR and the four-level ladder scheme is set to $6S_{1/2} \to 6P_{3/2} \to 47S_{1/2} \to 47P_{1/2}$ for a resonant millimeter-wave frequency $f_c = 37.406$ GHz \cite{anderson2020rydberg}. The specific parameter configurations of RAQR are obtained with the Alkali Rydberg Calculator (ARC) package \cite{vsibalic2017arc}. The proposed URformer is trained by the Adam optimizer with a cosine annealing learning rate scheduler over $T_\text{max} = 50$ epochs. Table I summarizes the main system parameters of the considered RAQR and training hyper-parameter setups of the proposed URformer. We compare the proposed URformer against several baseline methods: the classic GS and EM-GS algorithms, as well as two black-box deep learning approaches, i.e., the convolutional neural network (CNN) and the standard Transformer architecture \cite{b8}. For fair comparison, the number of iterations for the classic algorithms is set to be $T_\text{GS}=100$ to provide superior estimation accuracy for iterative algorithms. 

\begin{table}[t!]
\centering
\caption{ Parameter Setups for Channel Estimation in RAQRs}
\label{tab:params}
\begin{tabular}{l|c}
\hline
\textbf{Parameters} & \textbf{Values} \\
\hline
Number of users $K$ & 4 \\
Number of atomic antennas $M$ & 32 \\
Number of clusters $L$ & 4 \\
Number of subrays $C_l$ & 10 \\
DoA distribution $\theta_{l,c}^{(k)}$ &$\mathcal{U}(-\pi/2, \pi/2)$ \\
Number of URformer layers $T_\text{UR}$ & 10 \\
Number of Transformer encoders $L_\text{enc}$ & 3 \\
Embedding dimensions $d_{\text{model}}$ & 64 \\
Number of training samples & 20000 \\
Number of training batchsize & 32 \\
Initial learning rate & $1\times 10^{-3}$ \\
\hline
\end{tabular}
\end{table}

\begin{figure}[!t]
\centering
\includegraphics[width=2.5in]{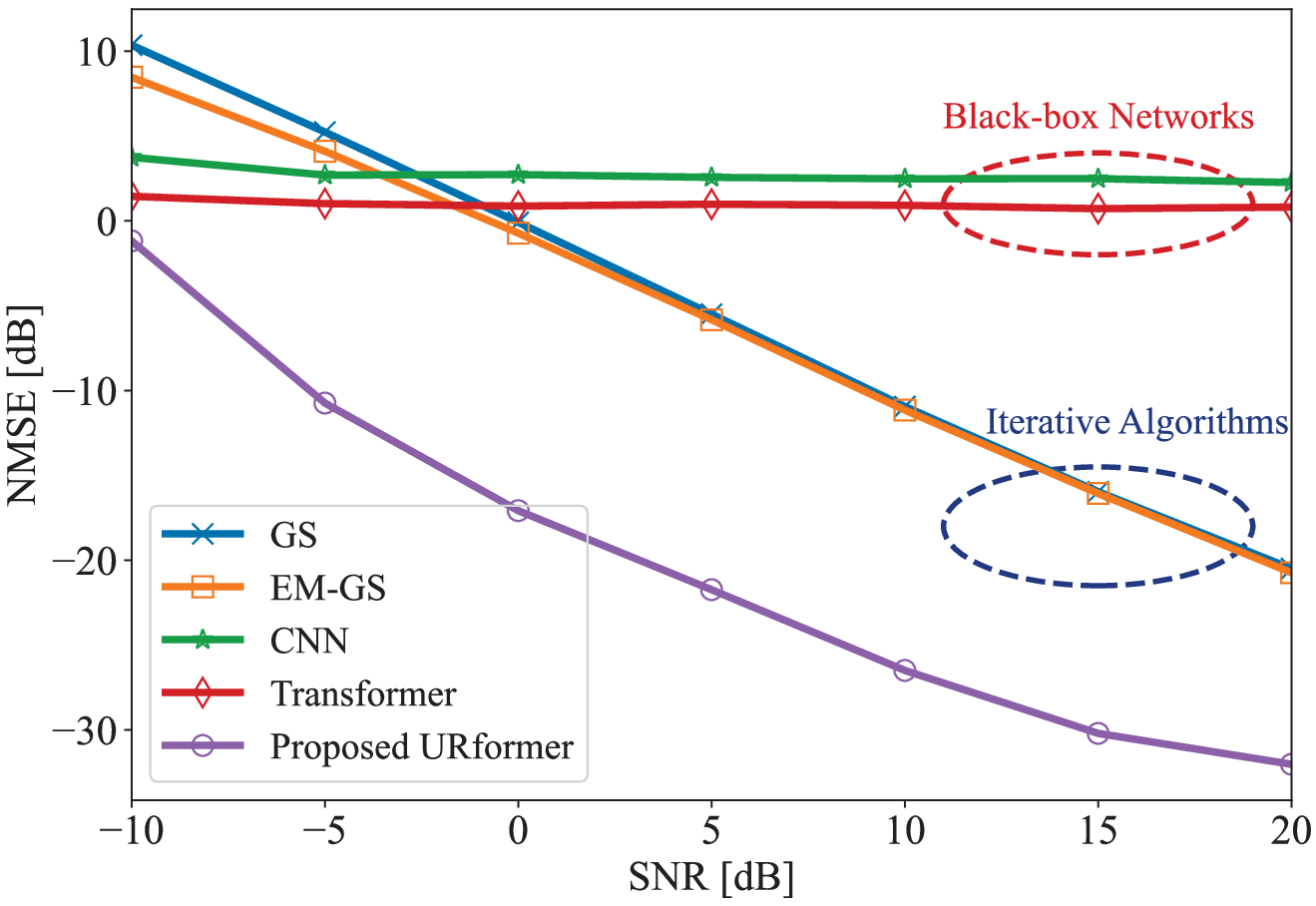}
 \caption{NMSE vs. SNR for different channel estimation schemes.}
\label{fig3}
\end{figure}
Fig.~\ref{fig3} illustrates the NMSE performance of all methods as a function of the received SNR that is defined as $\mathrm{SNR} = {\mathbb{E}\left(|\mathbf{H} \mathbf{s}_p|^2\right)}/{\mathbb{E}\left(|\mathbf{w}_p|^2\right)}$, where the reference-to-signal ratio (RSR) is set to $\mathrm{RSR} = {\mathbb{E}\left(|\mathbf{b}|^2\right)}/{\mathbb{E}\left(|\mathbf{H} \mathbf{s}_p|^2\right)} = 10$ dB and the number of pilots is set to $P=20$.
The results clearly demonstrate the superiority of the proposed URformer. In the challenging low-SNR regime, e.g., 0 dB, where the classic GS and EM-GS algorithms struggle significantly, URformer achieves a remarkable performance gain. This highlights the robustness of its learned modules, particularly the neural filter and Transformer block, which can effectively extract channel features even from very noisy measurements. In contrast, the black-box CNN and Transformer models fail completely, yielding a high error floor across all SNRs. This indicates their inability to solve the complex phase retrieval problem without the structural guidance of the underlying physical model. 

\begin{figure}[!t]
\centering
\includegraphics[width=2.5in]{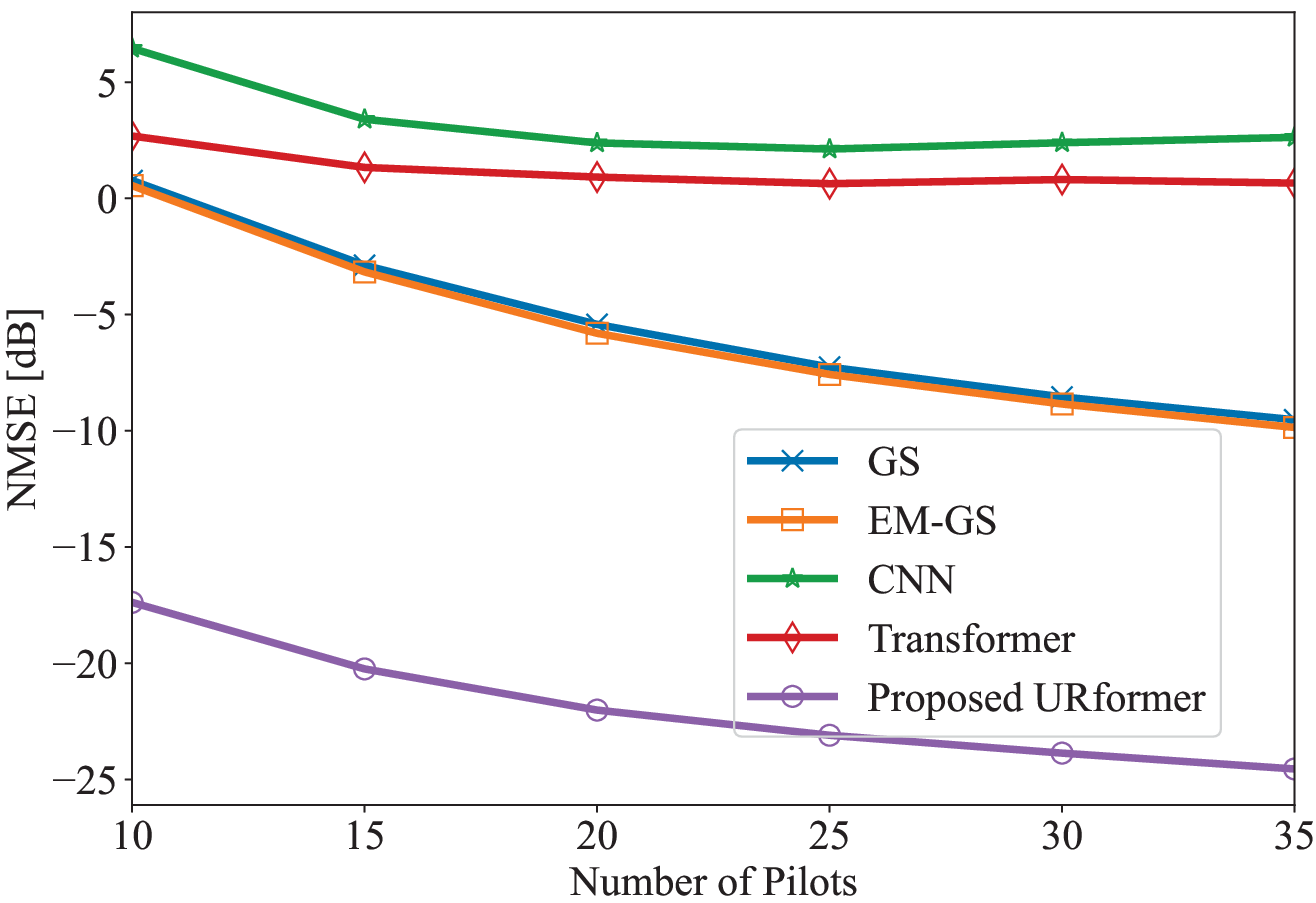}
 \caption{NMSE vs. number of pilots for different channel estimation schemes.}
\label{fig4}
\vspace{-5mm}
\end{figure}

Fig.~\ref{fig4} shows the NMSE performance versus the number of pilot transmissions $P$, evaluated at a fixed SNR of 5 dB. This analysis is crucial for assessing the pilot efficiency of each algorithm. As expected, the performance of all methods improves with an increased number of pilots. However, the proposed URformer exhibits a much steeper performance curve, indicating a significantly higher pilot efficiency. For instance, URformer achieves an NMSE of -20 dB with only $P=15$ pilots. This superior pilot efficiency allows for a reduction in pilot overhead, thereby increasing the spectral efficiency of RAQR-based communication systems.

\section{Conclusion}
In this letter, we investigated the channel estimation for RAQRs, which are characterized by a non-linear biased phase retrieval signal model from the measurement of holographic snapshots. Moving beyond the limitations of classic iterative algorithms and data-agnostic black-box models, we proposed a novel model-driven URformer architecture by unrolling a stabilized version of the classic EM-GS algorithm. Each unrolling layer of the proposed URformer is composed of a learnable neural filter, a stabilizing gate, and a channel Transformer module, allowing it to overcome the limitations of classic iterative methods. Numerical results demonstrated that the proposed URformer significantly outperformed the existing algorithms and showed substantially higher pilot efficiency. 

\bibliographystyle{IEEEtran}
\bibliography{IEEEabrv,refs_RAQR.bib}
\end{document}